\documentclass{WileyMSP-template}
\usepackage{textcomp}
\usepackage{xcolor}

\begin{document}
\pagestyle{fancy}
\rhead{\includegraphics[width=2.5cm]{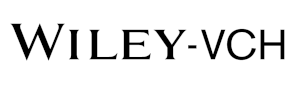}}

\title{Light-responsive nematic colloids and colloidal crystals}
%\title{Light activated defect-dynamics and structural transformation of nematic colloids}
\maketitle

% Author: Please give full first and last names for authors and include * after the name of all corresponding authors
\author{Devika Venkuzhy Sudhakaran,}
\author{Dinesh Kumar Sahu,}
\author{Osamu Haba,}
\author{Surajit Dhara*}

% Dedication
%\dedication{Optional dedication here. If no dedication is required, please leave blank}

% Affiliations: Please provide adacemic titles (Prof. or Dr.) for all authors where applicable, and include an institutional email address for all corresponding authors
\begin{affiliations}
Devika Venkuzhy Sudhakaran, Dr. Dinesh Kumar Sahu, Prof. Surajit Dhara \\
School of Physics, University of Hyderabad, Hyderabad - 500046, India\\
Email Address: surajit@uohyd.ac.in

Prof. Osamu Haba\\
Graduate School of Organic Materials Science, Yamagata University, Japan

\end{affiliations}

% Keywords: Please provide a minimum of three and a maximum of seven keywords, separated by commas


% Abstract should be written in the present tense and impersonal style (i.e., avoid we), and be at most 200 words long
\begin{abstract}
\begin{justify} 
Rational control over the periodic arrangement of particles by means of external stimuli is a technologically important aspect of colloidal science with important physical underpinnings. Here, a robust structural control of particle assemblies in a nematic liquid crystal (NLC) is demonstrated by dissolving trace amounts of light-responsive azo-dendrimer molecules which spontaneously get adsorbed on the particle surface. The azo-dendrimer molecules in the presence of external UV irradiation undergo conformational change (trans-cis); as a result, they transmit the mechanical torque to surrounding LC molecules and alter the near-field director orientation. The director re-orientation at the surface of the particles causes topological defect transformation which involves elastic dipoles, quadrupoles and hexadecapoles. The defect transformation can be emulated in colloidal assemblies towards different purposes such as rotation of chains and restructuring of 2D colloidal crystals. In this study, various topological aspects of light-activated defect transformation and its application in the collective manipulation of colloidal assemblies are presented. 
\end{justify}

\end{abstract}

% Text: Please use section headings and subheadings as specified below. For communications, all section headings apart from Experimental Section should be removed
% Please make the first reference to a display item bold: \textbf{Figure 1}
% Do not abbreviate Figure, Equation, etc.; display items are always singular, i.e., Figure 1 and 2.
% Equations are always singular, i.e., Equation 1 and 2, and should be inserted using the {equation} environment, not as graphics
% Please do not use footnotes in the text, additional information can be added to the Reference list.

\section{Introduction}
\begin{justify}
Stimuli-responsive structures formed by the self-assembly of micro- and nano-sized particles are indispensable tools for developing smart functional materials for emerging applications in the field of optics, photonics, microsensing, bio-interfaces etc~\cite{igor,stra,pine,dev,kang,chai,zerr,singh,zuhail,yuan,ivan}. A myriad of such geometrically diverse and dynamic structures could be conveniently assembled in an anisotropic liquid crystalline host matrix, owing to its long-range elasticity-mediated interactive forces. They have the ability to respond to external stimuli such as light, temperature, chemical substances and other external fields. For example, nematic liquid crystals (NLCs) exhibit directional elasticity and anisotropic electro-optic properties as a result of average orientation of the rod-shaped molecules along a preferred direction $(\hat{n})$, known as the director~\cite{genes,stark,igor1}. This gives rise to strong and long-range structural forces between particles suspended (nematic colloids) in NLCs. 
   Dispersion of foreign particles in an NLC creates orientational frustration in the director field, which results in singularities or topological defects at the LC-particle interface. The simplest case is that of spherical inclusions in NLCs, where the particles, depending on the surface anchoring conditions generally nucleate a point (hedgehog defect), ring defect (so-called Saturn ring) in the bulk, or a pair of antipodal surface defects (boojums)~\cite{ivan1,poulin,igor1,cherny,ivan2,cherny1}. The first two kinds of defects (point or ring defect in the bulk) appear while imposing homeotropic anchoring of LC on the particle's surface whereas the latter (boojum defects) is a consequence of planar anchoring of LC molecules at the particle surface. In the case of a hedgehog defect, the director symmetry and ensuing elasticity-mediated interactions are reminiscent of an electric dipole whereas, for a Saturn ring or Boojum defects, it resembles that of an electric quadrupole~\cite{igor,igor1,igor2,sriram,gama,igor3}. Hence, particles with hedgehog defects are also known as elastic dipoles; similarly, those with Saturn ring or boojum defects are called elastic quadrupoles. \\
   %################### 
\begin{figure}
  \includegraphics[scale = 0.55]{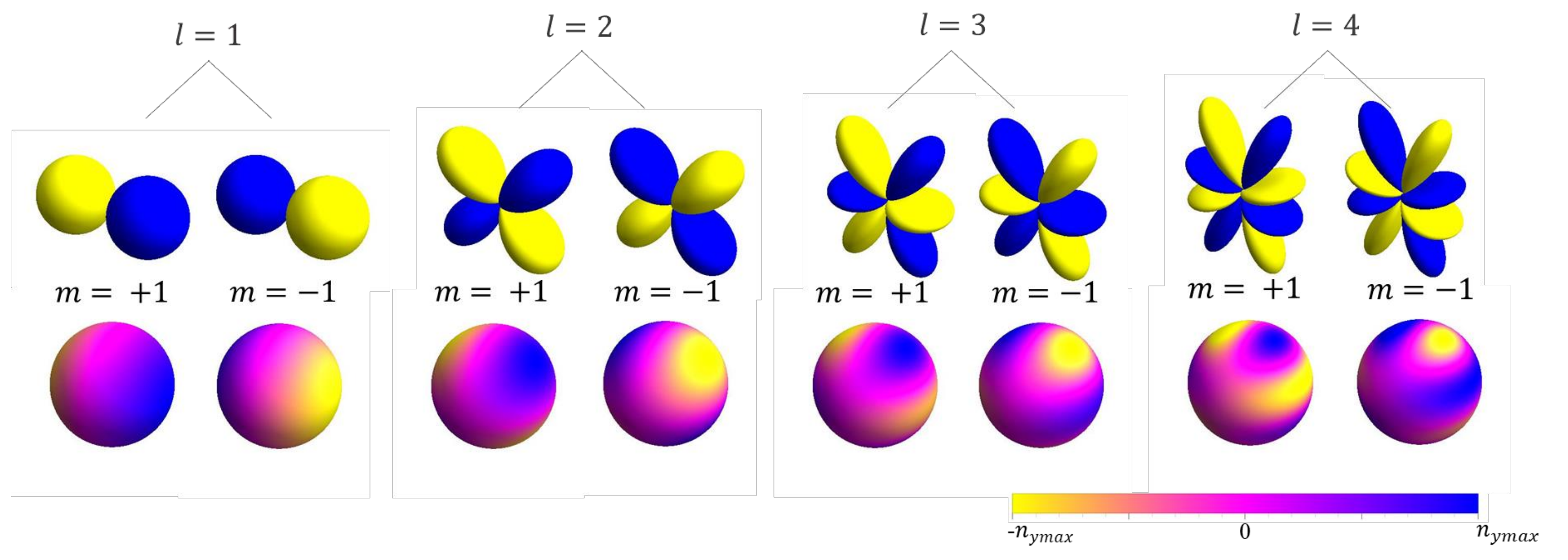}
  \centering
  \caption{Representation of elastic multipoles in terms of atomic orbitals. $s-$, $p-$, $d-$, and $f-$, atomic orbitals are drawn using spherical harmonics wave function; analogous elastic multipoles around a spherical particle with a colour-coded variation of $n_{y}$ are given below. }
  \label{fig:boat1}
\end{figure}
%################### 

Conferring desired anchoring of the nematic director at the LC-particle interface usually requires chemical functionalization of the surface such as pre-treatment with surfactants~\cite{align,dozov}. The high surface sensitivity of LCs (due to low interfacial energies) combined with emerging synthetic routes for developing surfaces with well-defined properties offers an attractive platform for the design of responsive soft materials. Here, we use an optically-triggered azo-dendrimer as an active interface to study the evolution of defect structures around microspheres and employ it in achieving structural transformation of 2D crystals and rotation of chains. The ability of the dendrimer to get adsorbed spontaneously on an ambient surface is useful in creating photo-sensitive surfaces of particles in the NLC \cite{seki,seki1,takezoe,takezoe1,takezoe2,merino}. It has been demonstrated that the azo-moieties, in response to highly localized UV light of low intensities, undergo reversible structural (cis-trans) isomerism, which drives collective re-orientation of LC molecules at the interface \cite{merino}. Under normal conditions, the azo-moieties adhere to the thermodynamically more stable trans-state, where LC molecules are held perpendicular to the surface. Upon illuminating with UV, they revert to the cis-state in which the LC molecules orient in a planar fashion. 
We show that the spatiotemporal director change forces a series of out-of-equilibrium defect structures around the microspheres during the process.
Such previously unanticipated defect states broaden the landscape of existing elastic interactions and ensuing self-assembled structures. As a direct application, we demonstrate the structural transformation of a 2D crystal of elastic dipoles when irradiated with UV light and measure substantial changes in the lattice parameters. The paper is organised as follows: First, we provide a brief introduction to the elastic multipoles and then present the defect dynamics of an isolated particle, followed by the rotation of a linear chain and restructuring of a 2D colloidal crystal.   
\end{justify}

\subsection{Results and Discussion}
\subsubsection{Colloidal dispersions as elastic multipoles}
\begin{justify}
Colloidal particles in a uniform nematic LC create near-field elastic distortion.
The far-field director $n_{0}$ (at length scales of several particle dimensions away from the surface) in the case of spherical particles has fewer distortions, hence the Euler-Lagrange equation can be linearized in this regime. Using the one-elastic-constant approximation (where different elastic constants yield similar values), the nematic LC free energy density takes the form~\cite{genes,senyuk,ivan3}:
     
\begin{equation}
f_{E}=\frac{1}{2}K \sum_{j=y,z}(\nabla n_{j})^2
\end{equation}
where, $K$ is the average Frank elastic constant, and $n \approx n (1,n_{y},n_{z})$ is the unit nematic director along the x-axis, and having perpendicular components $n_{j}(j = y,z \ll 1)$.
Minimization of the free energy density of (Equation 1) gives rise to the Laplace equation:
\begin{equation}
\nabla^{2} n_{j}=0,
\end{equation} 
the solution of which gives the multipole-expansion of the nematic director field as:

\begin{equation}
n_{j}(r,\theta,\phi)= \sum_{l=0}^{\infty} \sum_{m=-l}^{m=l} q_{lm}^{j}\frac{r_{0}^{l+1}}{r^{l+1}}Y_{l}^{m}(\theta,\phi)
\end{equation}
\noindent 
where $\theta,\phi$ represents the polar and azimuthal angles, respectively, $Y_{l}^{m}(\theta,\phi)$ are spherical harmonics,  $\textit{l}$ determines the order of multipole ($2^{l}$-pole), $-l \leq m \leq l$, $q_{lm}^{j}$ are dimensionless co-efficients of elastic multipoles, and $r_{0}$ is the radius of the particle.

%################### 
\begin{figure}
  \includegraphics[scale = 0.6]{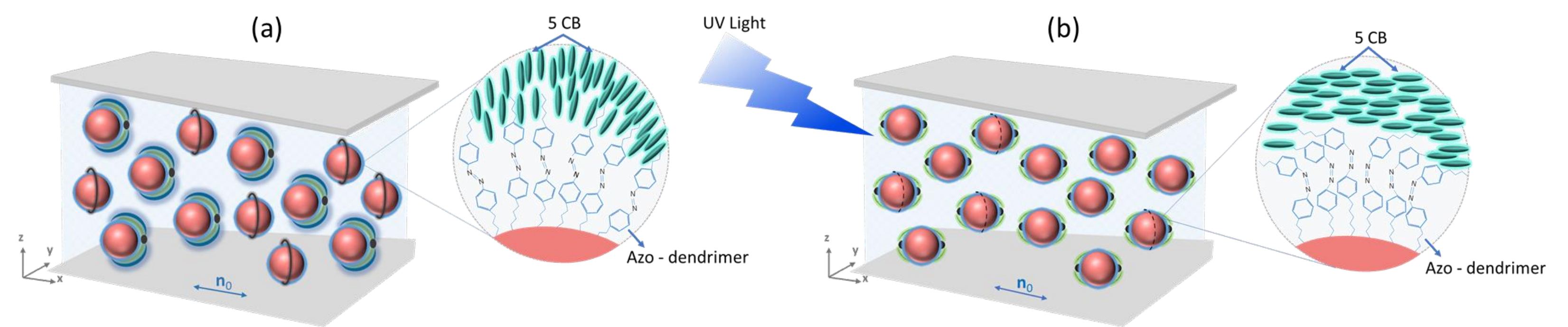}
  \caption{(a) Azo-dendrimer coated silica microspheres in a planar cell inducing hedgehog point or Saturn ring defects due to the $trans-$ conformation of the photoactive dendrimer favouring spontaneous homeotropic surface anchoring of LC molecules. The dendrimer-guided alignment of LC molecules is schematically shown in the zoomed view. (b) With UV illumination, the surface anchoring becomes planar as the dendrimer undergoes a conformational change to $cis-$ isomer state which results in boojum or hexadecapolar defects. Double-headed arrows represent the far-field nematic director $n_{0}$.}
  \label{fig:boat1}
\end{figure}
%#######################

In this form, it is imperative to draw an analogy of the nematic director field to the corresponding expansion of the electric field due to an electrostatic charge distribution \cite{genes,stark,igor1}. Thus, the nematic director $n_{j}(r,\theta,\phi)$ is expressed as a summation over elastic multipoles with each term in the expression (with co-efficient $q_{lm}^{j}$) representing a higher order multipole than the previous one (Figure 1). Here, it is to be noted that we have used spherical particles invariant to rotations about $\textit{x}-$ axis and have no azimuthal contribution to $\textit{n}(r)$. Therefore, the multipolar distortion induced by the particles is symmetric with respect to rotations about $n_{0}$. Consequently, the coefficient of the first term in Equation 3 (corresponding to elastic monopole) vanishes and the remaining higher order terms with $l > 0$ and $m = \pm 1 $ define the director distortions. Though higher-order terms are always present, the director symmetry of a particle-defect combination is always determined by the order of dominating multipole in the expression. \\

Most commonly observed are the elastic dipoles ($l = 1, m = \pm 1$) and elastic quadrupoles ($l = 2, m = \pm 1$) where the surface of the colloids induce either homeotropic or planar anchoring at the LC interface. Introducing particles of complex design may alter the structure of the defect (such as a distorted Saturn ring pinning to the sharp edges of a cube), but maintains the symmetry of distortion~\cite{dev}. Obtaining higher-order elastic multipoles like octupoles ($l=3$), hexadecapoles ($l=4$) etc has been a matter of challenge and significant interest~\cite{ivan2}. Recently, Senyuk $\textit{et al.}$, demonstrated that the continuous absorption or desorption of surfactants can drive the continuous transformation of elastic multipoles \cite{ivan3}. What we present here, is the light-assisted manipulation of elastic multipoles induced by spherical silica particles in a uniform nematic. Our experiments reveal the dynamics of the photo-reversible transformation of defects with varying symmetries and its effect on colloidal assemblies. 
\end{justify}   

\subsubsection{Light-activated defect dynamics around a spherical colloid}
\begin{justify}
The light-activated defect dynamics occurring at the LC-particle interface were studied using optically transparent silica microspheres (diameter 9.2 $\mu$m) dispersed in a nematic liquid crystal (5CB) mixed with $\approx 0.1$ wt\% azo-dendrimer (for details see experimental section). In the NLC, the dendrimer molecules get spontaneously adsorbed on the silica surface and act like a `command surface' which actively controls the LC orientation in response to external optical stimuli \cite{seki,seki1}. In the ground state, the dendrimer mostly prefers thermodynamically more stable $trans-$ conformation which prompts homeotropic (vertical) alignment of LC molecules. Hence, the microspheres induce either hedgehog point defects with dipolar symmetry $(l = 1)$ or Saturn ring disclinations with quadrupolar symmetry $(l = 2)$ (Figure 2a). Here, the surface nematic director \textbf{$n_{s}$} makes an angle $\alpha \approx 0$ (i.e., parallel) to the surface normal, \textbf{$\hat{s}$} (Figure 3b). When irradiated with UV light (wavelength $\approx$ 330 nm), the dendrimers undergo photo-isomerisation to cis-state which in turn rotates \textbf{$n_{s}$} by an angle of $90^{0}$, thereby rendering planar alignment ($\alpha \approx 90$) of the LC (Figure 2b). However, the change of surface anchoring from homeotropic to planar disturbs the existing balance between interfacial and bulk elastic free energies, thereby driving out-of-equilibrium defect transformations across the system to minimise the overall LC free energy. 
\end{justify} 

\subsubsection{Transformation from dipole to boojum}
\begin{justify}
%%%%%%%%%%%%%
\begin{figure}
  \includegraphics[scale = 0.6]{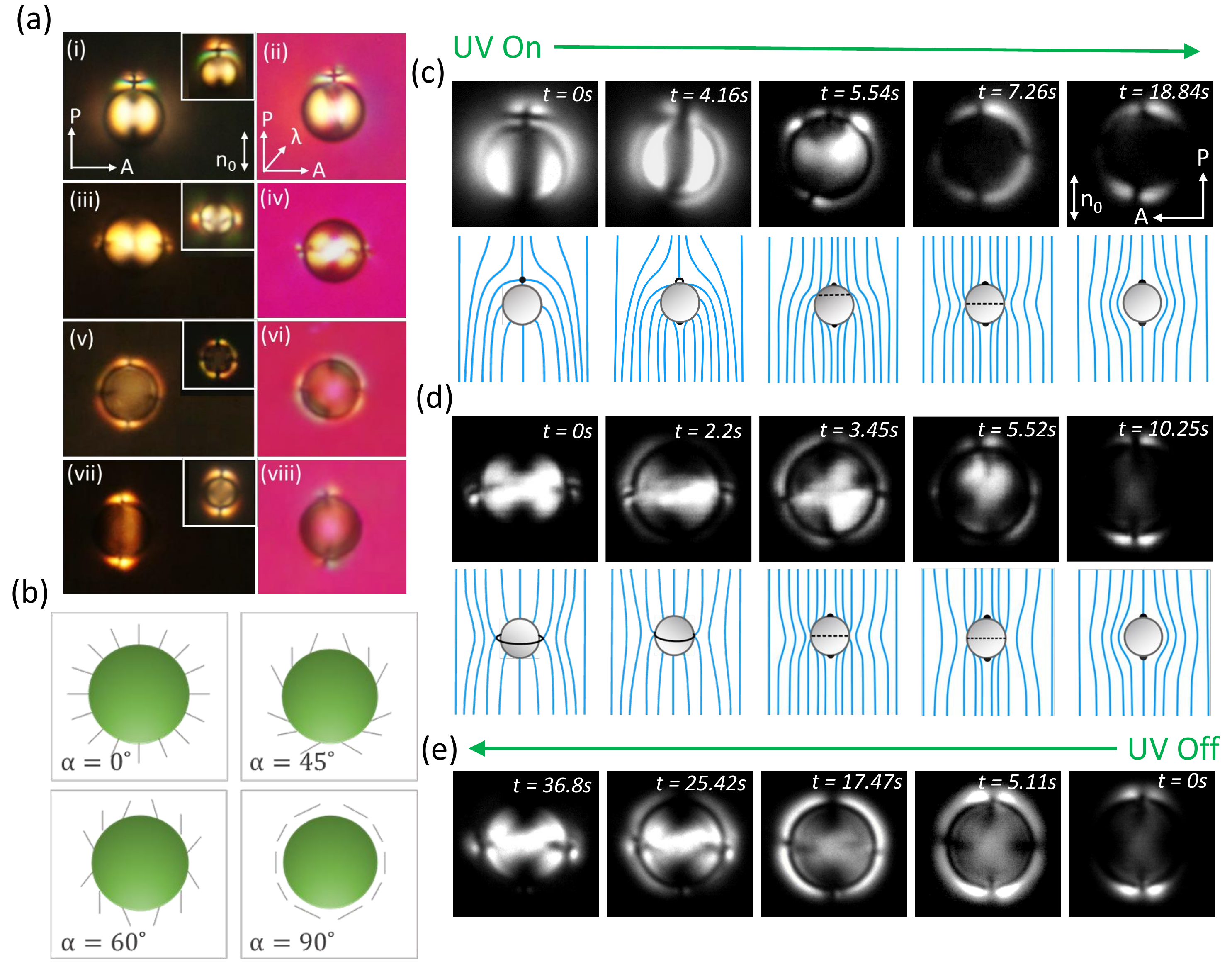}
  \centering
  \caption{(a) Polarising optical microscopy (POM) and corresponding $\lambda$ plate (inserted at $45^{0}$ between sample and analyser) images of (i-ii) hedgehog dipole, (iii-iv) Saturn ring quadrupole, (v-vi) elastic hexadecapole, and (vii-viii) boojum quadrupole. Particle size - 9.2 $\mu$m. POM images of 5.24 $\mu$m particles are shown in the inset. (b) Schematics of surface director tilt for different values of the polar anchoring angle, $\alpha$. Transformations of elastic multipoles from (c) a hedgehog dipole to a boojum quadrupole (Movie-1), and (d) a Saturn ring quadrupole to a boojum quadrupole (Movie-3) under UV irradiation. Corresponding schematics of the director profile around the particle are shown below. (e) Reverse transformation to a Saturn ring quadrupole when UV is switched off (Movie-2 and 4). P, A represents crossed polariser and analyser, the far-field director is given by n$_{0}.$}
  \label{fig:boat1}
\end{figure}
%%%%%%%%%%%%%%

The light-driven spatiotemporal evolution of defect structure around a microsphere with initial hedgehog point defect in the bulk is shown in Figure 3c. As can be seen, the point defect in the bulk evolves into a series of complex defect structures before stabilising the final quadrupolar state (a pair of antipodal boojums). It is evident that these complex structures appear due to the increase in the tilt angle ($\alpha$) of the director with time.
 Such transformation of defects from elastic dipole to higher order multipoles has been recently proposed by Zhou \textit{et al.}~\cite{ivan4} in nematic colloids with conic degenerate anchoring, where $\alpha$ varies from 0 to $90^{\circ}$ (Figure 3b). In our experiment, the different stages of evolution of a dipolar colloid when excited by UV irradiation, captured at different time intervals, reveal similar director distortions (Figure 3c, Movie-1\cite{sup}) as predicted. Following it, we present a qualitative discussion of the defect transformation. \\
 
Initially, we consider an elastic dipole (Figure 3c [$t=0$ s]), i.e., a particle with an accompanying hedgehog point defect of charge -1 ($l=1, m=1$). Upon illumination with UV, the tilt angle $\alpha$ increases and we observe what is called a `CA dipole': a texture with a hedgehog point defect at one pole of the particle and a boojum defect at the opposite pole (Figure 3c [$t=4.16$ s] and schematic). It may look similar to the commonly observed homeotropic elastic dipole, but a closer observation reveals the presence of a dark spot (the boojum) at the opposite pole, surrounded by two weak bright lobes. The theoretical model suggests that at $\alpha = 45^{\circ}$, the 2D topological charge -1 splits unequally between the upper ($-3/4$) and lower ($-1/4$) surface defects (see schematic), in compliance with the conservation of topological charges~\cite{genes,stark,igor1}. With further increase in the value of tilt angle, a surface ring nucleates at the upper pole (Figure 3c [$t=5.54$ s]) and starts migrating downwards (Figure 3c [$t=7.26$ s]). Both these structures composed of a pair of boojum and Saturn ring quadrupoles are characteristic of an elastic hexadecapole ($l=4, m=1$), reported recently \cite{ivan4}. The POM micrograph of a hexadecapole shows eight bright lobes: two each at the poles and the equator, separated by eight dark regions (see inset of Figure 3a (v)). The simulation predicts that the hexadecapolar structure can be stabilised over a wide range of tilt angles ($40^{\circ} \leq \alpha \leq 60^{\circ}$), and the corresponding distortions could be distinguished by the relative brightness of their lobes~\cite{ivan4}. As shown in Figure 3c [$t=7.26$ s], just before its transition to antipodal boojum, we observe a distortion pattern similar to $\alpha \approx 60^{\circ}$, where the net topological charge -1 is equally split (-1/3) among the upper and lower surface point defects, surface ring defect at the equator. Finally, when $\alpha$ increases to $90^{\circ}$, a pair of antipodal boojum ($l=2, m=1$) of strength $-1/2$ each is formed (Figure 3c [$t=18.84$s]).   
\end{justify}   

\subsubsection{Transformation from Saturn ring to boojum}
\begin{justify}
Other than hedgehog point defects, particles with homeotropic anchoring also form Saturn ring defects (topological charge = -1) with quadrupolar symmetry ($l=2, m=-1)$ (Figure 3a(iii,iv)). The director at the surface of these particles is initially oriented hometropically and gradually undergoes tilting (conic anchoring) when the UV light is switched on (Figure 3d, Movie-3 \cite{sup}). With increasing tilt angle, the Saturn ring vanishes and a surface ring defect (charge = -1/3) along with a pair of surface point defects (charge = -1/3, each) emerge at the poles (Figure 3d [$t=3.45$ s]). That means, the elastic quadrupole evolves to a higher-order hexadecapole ($l=4, m=1$)(Figure 3a(v,vi)). Further, with an increasing value of $\alpha$, the surface ring defects diminish and ultimately transform into a pair of antipodal boojum defects (charge = -1/2, each) prevailing tangential anchoring.

Whether we start with a point defect (dipolar configuration) or Saturn ring, the end result is always antipodal boojums upon UV irradiation. When the UV light is turned off, the boojums, in both cases, retrace the path to a bulk Saturn ring via an intermediate elastic hexadecapole as shown in Figure 3e (Movie-2 and 4 \cite{sup}). It is to be noted that the intensity of light significantly affects the final state attained. For example, we often observed that low light intensities ($<45 \mu$W/mm\textsuperscript{2}) caused the defect evolution in an elastic dipole to stop at any of the intermediate states, where $0\leq\alpha \leq 90$. Thus, UV light of controllable brightness offers a reliable route for stabilising transient topological defects which otherwise, are rarely observed.
%%%%%%%%%%%%%
\begin{figure}
  \includegraphics[scale = 0.5]{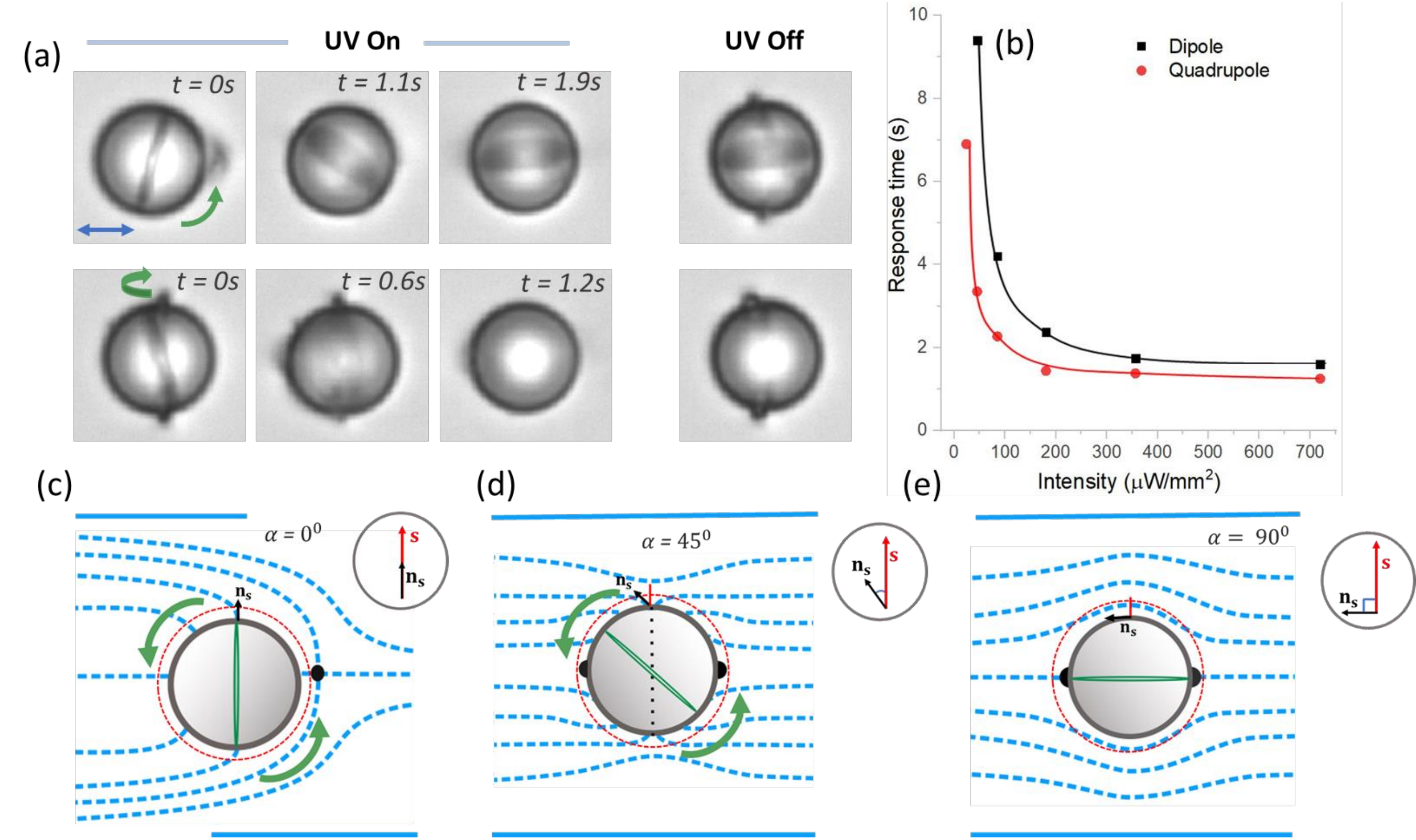}
  \centering
  \caption{(a) Snapshots depicting the rotation of a sphere (in-plane or out-of-plane) by almost $90^{0}$ after the UV light is switched on (Movie-5). (b) Comparison of the response time (in seconds) versus UV intensity ($\mu$W/mm\textsuperscript{2}) for elastic dipoles and quadrupoles. The solid curves are guidelines for the eye. (c-e) Schematics of the director field around a sphere for different values of $\alpha$. Green curved arrows represent the sense of rotation of the particle.}
  \label{fig:boat1}
\end{figure}
%%%%%%%%%%%%%%

%%%%%%%%%%%%%%%
\begin{figure}
  \includegraphics[scale = 0.5]{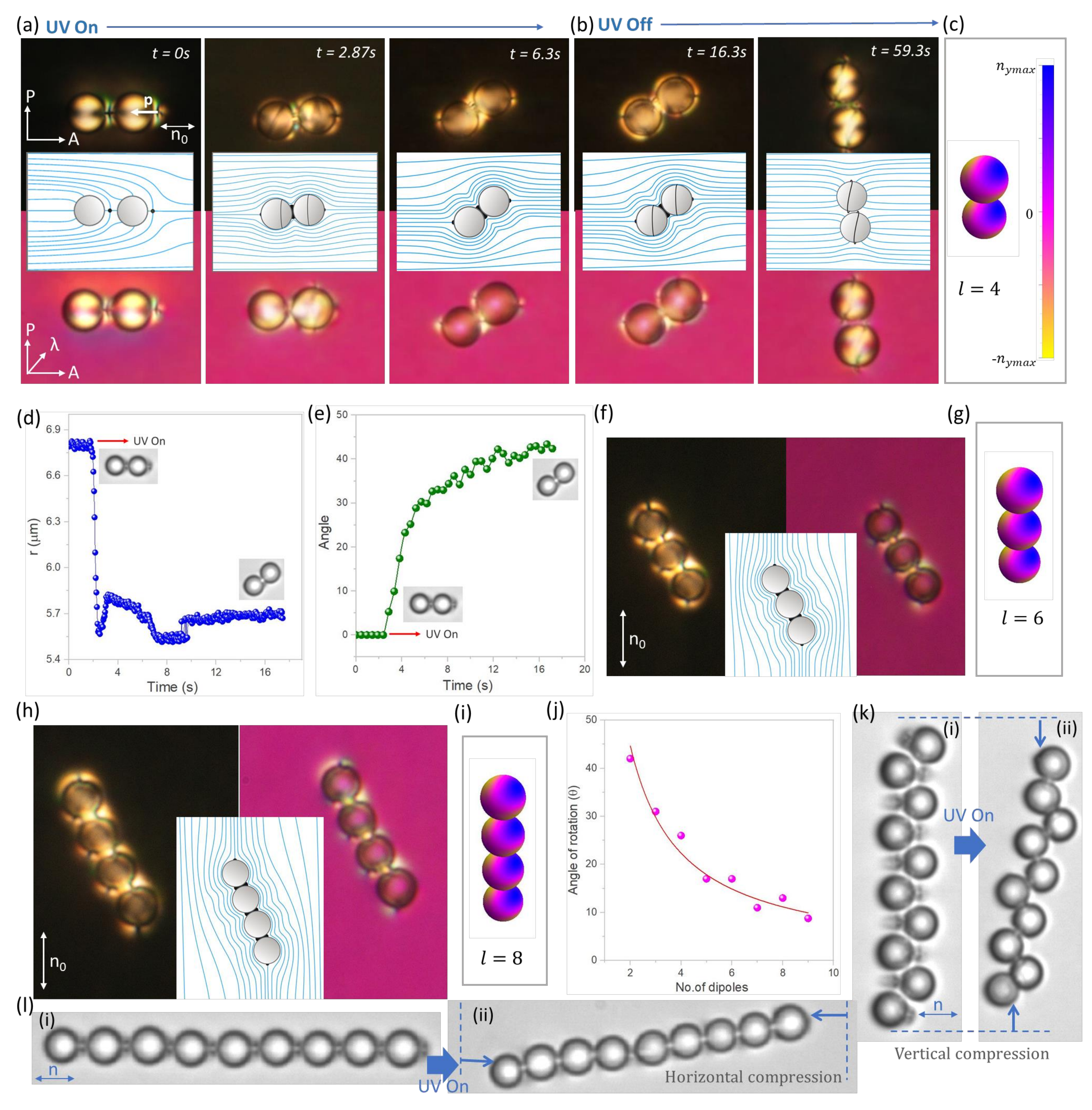}
  \centering
  \caption{ (a) Polarising microscopy (POM) images with (second row) and without (first row) a $\lambda-$ plate during the transformation of a pair of self-assembled hedgehog dipoles to boojum quadrupoles in the presence of UV. Schematics of the surrounding director field are shown in the inset. (b) POM images with (second row) and without (first row) a $\lambda-$ plate (director field schematic in the inset) during the transformation to Saturn ring quadrupoles when UV is switched off. (c) Colour maps of director distortion at the surface of the particles in the final equilibrium state resembling an elastic multipole with $l = 4$. The colour bar on the right represents the normalised $y-$ component of the nematic director. (d)The temporal variation of interparticle separation (r, length of the line joining the centre of spheres) with UV illumination. (e) Corresponding angle of rotation $\theta$ of the chain with respect to n$_{0}$. (f) POM and corresponding $\lambda-$ plate image of the final equilibrium state of a chain of three particles (Schematic of the surrounding director profile is shown in the inset). (g) Colour maps of the director profile at the surface of the particles resembling an elastic multipole with $l = 6$. (h)  POM and corresponding $\lambda-$ plate images of the final equilibrium state of a  chain of four particles (Movie-6). (i) Colour maps of the director profile at the surface of the particles resembling an elastic multipole with $l = 8$. (j) Variation of the maximum rotation angle  $\theta$ with an increase in chain length, for a fixed intensity of UV. (k) A vertical chain of dipoles undergoing compression in the $y-$ direction. (l) A horizontal chain of dipoles undergoing compression in the $x-$ direction. P, A represents the polariser and analyser. The double-headed arrow denotes the nematic director, n$_{0}$.} 
  \label{fig:boat1}
\end{figure}
\end{justify}
%%%%%%%%%%%%%%%%

\subsubsection{Rotation of the particles during defect-transformation}
\begin{justify}
It is observed that the light-activated defect dynamics is accompanied by the rotation of the particles as shown in Figure 4a (Movie-5 \cite{sup}). When irradiated with UV, the orientation of the director at the surface changes from homeotropic to planar and simultaneously the sphere rotates by almost $90^{\circ}$ either in-plane (upper row of Figure 4a) or out-of-plane (bottom row of Figure 4a). A schematic diagram of the corresponding director structure is shown in Figure 4(c-e). 
The director rotation gives rise to elastic torque which is proportional to $KD$, where $K$ is the average Frank elastic constant and $D$ is the diameter of the particle. This torque is counteracted by a viscous torque which is proportional to $\eta D^{3}$, where $\eta$ is the mean viscosity of the LC~\cite{yuan,takezoe1}. Considering the anchoring is stronger for the planar director (the case with higher intensity) \cite{takezoe1} switching response time of the sphere can be expressed as $\tau=\eta D^{2}/K$. Taking $\eta=75$ mPa s and $K=5$ pN, the calculated response time $\tau\simeq1.5$ s,  which agrees well with the experimentally observed response time (Figure 4b). Low UV intensity corresponds to weak surface anchoring \cite{takezoe1}, where the elastic torque exerted by the director depends on a characteristic length $D+\xi$, where $\xi$ is the penetration length given by $\xi=K/W$. In this case, the response time is given by $\tau=\eta(D+\xi)^{2}/K\simeq6$s which is also in good agreement with experiments. A comparison of the response time of elastic dipoles and quadrupoles for a given intensity of UV always shows higher values for dipoles (Figure 4b). This behaviour is expected, as the transformation from dipolar to boojum configuration involves more intermediate defect states than quadrupole to boojum. 
\end{justify}

\subsubsection{Rotation of colloidal chains}
\begin{justify}
The response of a dimer consisting of two colinear elastic dipoles with negative elastic dipole moment, \textbf{p$<0$} (directed along the -$x$ axis) under UV light is shown in Figure 5(a,b). As a result of the director re-orientation occurring at the surface of each dipole, they undergo defect transformation and rotate collectively nearly by $45^{\circ}$ in the anticlockwise direction as shown in Figure 5a. In particular, under UV illumination, the surface anchoring of the dimers changes from homeotropic to conic degenerate ($0^{\circ}\leq\alpha\leq90^{\circ}$) state, where the hedgehog point defects transform to a combination of boojum and neck defects~\cite{senyuk} as illustrated in the schematic. Here, the elastic distortion of the dimers corresponds to that of an elastic multipole of order $l = 4$ (Figure 5c). Changing the polarity of dipole moment (\textbf{-p}) reverses the sense of rotation of the dimers. 

Switching off UV ( Figure 5b) again causes the director to adopt homeotropic alignment, hence driving defect transformation with an intermediate state having a combination of Saturn ring, neck and boojum defects (Figure 5b [$t=10$ s] and underneath schematic), which later transforms to Saturn rings (Figure 5b [$t=53$ s]). The rotation angle further increases during this transformation. The temporal variation in inter-particle separation (distance between the centre of two spheres, $r$) and the angular tilt with respect to the far field director $n_{0}$ for a typical pair of particles is plotted in Figure 5(d,e), which shows a considerable reduction ($\approx$1$\mu$m) of the interparticle separation. The effect of UV on larger chains is studied systematically by adding more particles to it with the help of laser tweezers. A  chain of three dipoles under UV irradiation transforms to a final state as shown in Figure 5f (schematic in the inset) whose director structure resembles that of an elastic multipole of order, $l=6$ (Figure 5g). Similarly, a chain of four dipoles transforms to an elastic multipole of higher order, $l=8$ (Figure 5(h,i) Movie-6 \cite{sup}). Thus, simply by adding more particles to the chain, we can create higher-order elastic multipoles. It is apparent that with increasing chain length,  the angle of rotation with respect to the director, decreases ( Figure 5j). This could be rationalised by drawing an analogy with the rotation of a rod with increasing length in a nematic LC. The elastic free energy associated with the rotation of a rod of length $L$  at an angle $\theta$ with respect to the director is given by: $F_{el} = 2\pi CK \theta^{2}$
where, $C = \frac{L}{2}log(L/d)$ \cite{genes1}. At equilibrium, the elastic torque $\tau_{el} =-\frac{\partial F}{\partial\theta}\propto L \theta$ is balanced by the viscous torque $\tau_{vis} \propto \eta S^{3} $, where $S$ is the gap between the upper and lower glass plates. Hence, the angular tilt $\theta \propto 1/L$, as shown in Figure 5j. 
Figures 5(k,l) show two kinds of dipolar chains assembled perpendicular and parallel to $n_{0}$ making use of the anisotropic interactions specific to LCs. In a perpendicular chain, any two successive dipoles are oriented oppositely and in parallel chains they are colinear. As evident from the figure, the defect transformations not only cause rotation, but also reduction in length either along $x-$ (for parallel chains) or $y-$(for perpendicular chains) direction. It immediately comes to notice that both kinds of length compression can be employed in a 2D crystal structure obtained by stacking several such dipolar chains, thus allowing robust control over structural transformations.

%%%%%%%%%%%%%%%%%
\begin{figure}
  \includegraphics[scale = 0.6]{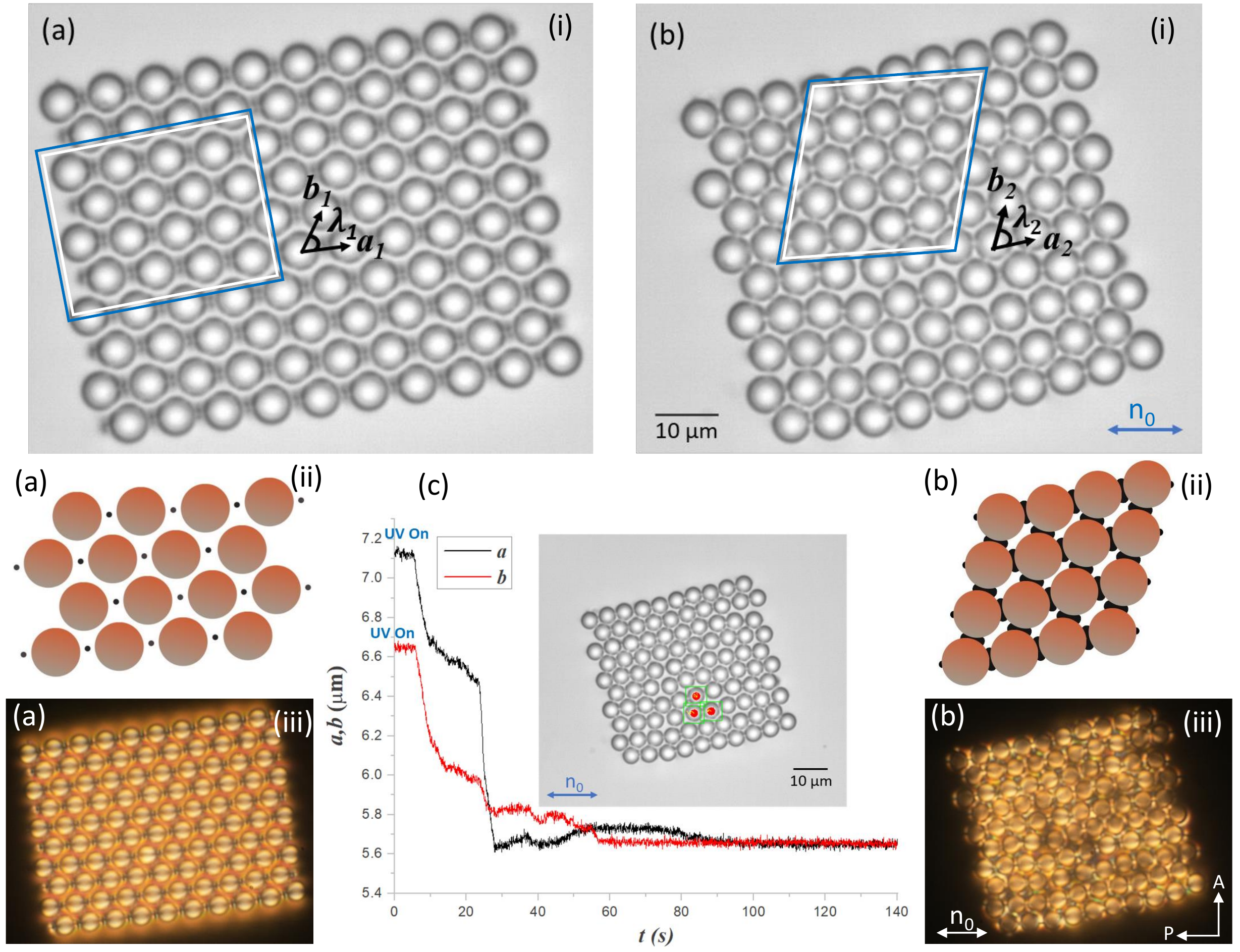}
  \centering
  \caption{(a-i) A 2D crystal of elastic dipoles formed by directed assembly using the laser tweezers. Equilibrium lattice parameters: $a_{1}$ = 6.67 $\mu$ m, $b_{1}$ = 6.11 $\mu$ m, $\lambda_{1}=54.4^{0}$, Area $\approx$ 3419 $\mu m^{2}$. (a-ii) Schematic of a $4 \times 4$ dipolar crystal (area enclosed by the rectangular selection). (a-iii) Polarising microscopy texture. (b-i)  Crystal after transformation by UV irradiation. Lattice parameters: $a_{2}$ = 5.42 $\mu$m, $b_{2}$ = 5.30 $\mu$ m, $\lambda_{2} = 63.6^{0}$, Area $\approx$ 2681 $\mu m^{2}$. (b-ii) Schematic of a $4 \times 4$ crystal after transformation (area enclosed by the rectangular selection). (b-iii) Polarising microscopy texture. (c) Change in lattice parameters during crystal transformation (Movie-7). Particles tracked for time-dependent lattice parameters are marked red. P, A represents polariser and analyser. Double-headed arrow denotes the far-field director. }
  \label{fig:boat1}
\end{figure}
\end{justify}
%%%%%%%%%%%%%%%%%

\subsubsection{Structural transformation of a 2D colloidal crystal}
\begin{justify}
Using a dynamic laser tweezers setup, the elastic dipoles were directed to form a regular 2D colloidal crystal as shown in  Figure 6a(i-iii). Here, the anchoring mismatch at the cell boundary (planar anchoring) and the surface of the spheres (homeotropic anchoring) introduces a wall-dipole repulsion which balances the gravitational pull and causes the particle to levitate in the LC channel \cite{oleg}. The structural change of the 2D crystal after UV radiation is shown in  Figure 6b(i-iii).
 Analysis of the structure reveals that before UV irradiation, the particles form a unit cell of hexagonal lattice ($a \neq b$) with average lattice parameters $a_{1}=6.33~\mu$m and $b_{1}=5.75~\mu$m and an angle, $\lambda \approx54.4^{\circ}$ between them. The temporal changes of the lattice parameters are shown in Figure 6c which shows that within about 30 s the transformation is completed and eventually $a=b$ at a longer time (Movie-7 \cite{sup}). After irradiation with UV, the surface director $n_{s}$ of the elastic dipoles tend to rotate by an angle of $90^{\circ}$, which is not possible now as the free rotation of the nematic director is arrested due to additional elastic constraints imposed by the neighbouring particles. This results in a conic degenerate anchoring of $n_{s}$, with a topological defect configuration consisting of neck defects at particle joints and boojum defects at the ends of each chain (figure 6(b-ii)). Hence, they form an oblique square lattice with ($a\approx b$) and lattice parameters, $a_{2}=5.58~\mu$m, $b_{2}=5.67~\mu$m, and $\lambda = 63.6^{\circ}$. An overall compression in the crystal area by about 24\% of the initial area is measured.
  
\end{justify}
 
\section{Conclusion}
\begin{justify}

Our experiments demonstrate light-activated defect-dynamics occurring at the interface of spherical inclusions embedded in an anisotropic nematic liquid crystalline matrix and its usefulness towards realising structural transformations across 2D colloidal assemblies. This is achieved by decorating the surface of microspheres with azo-dendrimers exhibiting conformational change ($cis-trans$) when perturbed with UV irradiation of sufficient intensity. The UV irradiation triggers particle rotation and drives out-of-equilibrium defect transformations in the system where the evolving director field reveals higher-order elastic multipoles. These intermediate defect states of higher order and complexity is pivotal in exploring the richness of anisotropic interactions in LCs and their potential in building remotely reconfigurable soft crystals. As a direct application, the defect-transformation-induced restructuring of a 2D colloidal crystal is demonstrated. Our findings can be extended to nematic emulsions as well as having wide-ranging applications in microfluidics. 
\end{justify}

\section{Experimental Section}
\begin{justify}
\textit{We prepared an azo-dendrimer-NLC mixture by dissolving a small quantity ( $<0.1$ wt\%) of poly (propyleneimine) based azo dendrimer in pentyl cyanobiphenyl (5CB) liquid crystal (Sigma Aldrich)~\cite{seki2}. The azo-dendrimer has an absorption peak at 360 nm of incident light, where it reconfigures to the isomeric $cis-$ state \cite{seki2}. Initially, the  dendrimer was dissolved in chloroform solvent and then stirred continuosly for 24 hrs using a magnetic stirrer. Any clusters were removed by filtering the solution through a micro seive of diameter 0.22 $\mu$m. It was then mixed with 5CB after which the chloroform solvent is allowed to evaporate completely at room temperature under a vacuum. Silica microparticles of diameter 5.24 and 9.2$\mu$m (Bangs Laboratory, USA) were dispersed into the as-prepared azo dendrimer-5CB mixture and sonicated for 10 mins. to obtain uniform azo-dendrimer coating on the silica surface. Planar cells conferring uni-directional alignment of the nematic director were fabricated using the following technique: ITO-coated glass plates washed thoroughly and dried under nitrogen atmosphere are spin-coated with a polyimide (AL-1254) and annealed at $180 ^{0}$C for 1 hr. The polyimide-treated glass plates are then rubbed using a velvet cloth along a specific direction.  Two glass substrates having opposite rubbing directions are attached with silica spacers and a UV-curable optical adhesive to form a capillary-like arrangement. }\\
\indent\textit{For observations, an inverted polarizing optical microscope (Nikon eclipse Ti-U) with a halogen lamp (operating at 12V, 100W) as incident light source and 60X water immersion objective (Nikon, NIR Apo 60/1.0) was used. The director distribution at the particle surface was mapped with the help of a full retardation wave plate (530 nm) placed at $ 45 ^{0}$ with respect to the analyser. A fluorescent source (Nikon, Intensilight C - HGFI) operating on a mercury lamp with controllable brightness connected to the microscope provided UV light. For this, light from the fluorescent source was passed through a UV filter cube (330 - 380 nm). An optical tweezer built on the inverted microscope employing a laser light of 1064 nm and controlled by an acousto-optic deflector, helped in setting traps and manipulating particles. Image acquisition and video recording were performed by a CCD camera and a Nikon DS-Ri2 colour camera attached to the microscope. Data analysis was performed using ImageJ software.} \\
\end{justify}

\medskip
\textbf{Supporting Information} \par %Please delete the Suppporting Information statement if it is not applicable. Please supply Supporting Information in another file. Supporting information should not be provided in .tex format
Supporting Information is available from the Wiley Online Library or from the author.

% Acknowledgements
\medskip
\textbf{Acknowledgements} \par %delete if not applicable))
DVS and DKS acknowledges DST, Government of India for INSPIRE fellowship. SD acknowledges support from IoE (UoH/IoE/RC1-20-010).

%\textbf{References}\\
% References
\medskip

% Use the following code if you wish to generate your bibliography with BibTeX;
% replace the string "MSP-template" below with the name(s) of
% the BibTeX data base(s) you want to use.
% The resulting bibliography-output (the content of the .bbl file)
% must be pasted back into this file before submission.
% Please also include your BibTeX data base file(s) in your submission
% so that we can re-run BibTeX if necessary.
%
%\bibliographystyle{MSP}
%\bibliography{MSP-template}

%1	((Journal articles)) a) A. B. Author 1, C. D. Author 2, Adv. Mater. 2006, 18, 1; b) A. Author 1, B. Author 2, Adv. Funct. Mater. 2006, 16, 1.\\
%2	((Work accepted)) A. B. Author 1, C. D. Author 2, Macromol. Rapid Commun., DOI: 10.1002/marc.DOI.\\
%3	((Books)) H. R. Allcock, Introduction to Materials Chemistry, Wiley, Hoboken, NJ, USA 2008.\\
%4	((Edited books or proceedings volumes)) J. W. Grate, G. C. Frye, in Sensors Update, Vol. 2 (Eds: H. Baltes, W. GÃ¶pel, J. Hesse), Wiley-VCH, Weinheim, Germany 1996, Ch. 2.\\
%5	((Presentation at a conference, proceeding not published)) Author, presented at Abbrev. Conf. Title, Location of Conference, Date of Conference ((Month, Year)).\\
%6	((Thesis)) Author, Degree Thesis, University (location if not obvious), Month, Year.\\
%7	((Patents)) a) A. B. Author 1, C. D. Author 2 (Company), Country Patent Number, Year; b) W. Lehmann, H. Rinke (Bayer AG) Ger. 838217, 1952.\\
%8	((Website)) Author, Short description or title, URL, accessed: Month, Year.\\
%9	âŠ((Please include all authors, and do not use âet al.â))

\end{document}